# Globalisation and the Decoupling of Inflation from Domestic Labour Costs

E. Kohlscheen and R. Moessner[1,2]

**Abstract**

We provide novel systematic cross-country evidence that the link between domestic labour markets and CPI inflation has weakened considerably in advanced economies during recent decades. The central estimate is that the short-run pass-through from domestic labour cost changes to core CPI inflation decreased from 0.25 in the 1980s to just 0.02 in the 2010s, while the long-run pass-through fell from 0.36 to 0.03. We show that the timing of the collapse in the pass-through coincides with a steep increase in import penetration from a group of 10 major manufacturing EMEs around the turn of the millennium – which signals increased competition and market contestability. Besides the extent of trade openness, we show that the intensity of the pass-through also depends in a non-linear way on the average level of inflation.

[1] Both authors are Senior Economists at the Bank for International Settlements. Centralbahnplatz 2, 4002 Basel, Switzerland. *Email addresses*: emanuel.kohlscheen@bis.org, richhild.moessner@bis.org.
[2] We are grateful for comments from Markus Eberhardt and to Emese Kuruc for excellent help with the data. The views expressed in this paper are those of the authors and do not necessarily reflect those of the Bank for International Settlements.

# 1. Introduction

Labour market developments are a central input for monetary policy decisions. In particular, wage pressures and the extent to which these could translate into more generalised price increases are typically monitored on a continuous basis. This is partly because wage–price spirals have historically led to adverse inflationary outcomes.

In this paper we focus on the link between domestic labour markets and CPI inflation, as well as the main factors that affect it. We provide novel cross-country evidence on the evolution of the pass-through of variations in domestic labour costs to inflation. We do so by resorting to dynamic panel estimations which allow for heterogeneity in coefficients across countries. We then test possible drivers that are behind the strength/weakness of this link using econometrics as well as non-parametric machine learning.

Overall, we find that the link between domestic unit labour cost growth and inflation has weakened considerably in advanced economies during recent decades. The central estimate is that the short-run pass-through from domestic labour cost variations to core CPI inflation decreased from a significant value of 0.25 in the 1980s to an insignificant value of just 0.02 in the 2010s. Similarly, the long-run pass-through fell from 0.36 in the 1980s to an insignificant value of 0.03 in the 2010s. We show that the timing of the collapse in the pass-through coincides with a steep increase in import penetration from a group of major manufacturing EMEs around the turn of the millennium, which implied increased competition and market contestability for advanced economies.

Overall, globalisation and trade openness have been the dominant factors behind the relative decoupling, while lower inflation levels have also played a role to lower domestic ULC–CPI pass-throughs. Using an agnostic machine learning technique (a random forest, based on 1,000 regression trees), we conclude that the extent of pass-through increases non-linearly with the level of average inflation. While the pass-through is generally very low for open economies with modest levels of inflation, it tends to rise rapidly when average inflation moves above 2%.

**Relation to the literature.** Our contribution to the literature is threefold: first, we establish and time the systematic decline of the pass-through across a broad set of countries; second, we show how the decline is linked with globalization, and establish the precise channel through which globalization matters (namely, through imports from lower-wage countries); third, we are able to identify the precise way in which the average level of inflation also feeds into the pass-through (through econometrics and machine learning, which can easily accommodate non-linearities).

Our article follows the spirit of Auer *et al* (2013), who show that import competition from low-wage emerging economies strongly reduced producer prices in a number of European countries. Several papers have continued to explore this link since then. Typically, in contrast to ours however, they have focused on individual economies. For instance, Peneva and Rudd (2017) find that the pass-through of labour costs to prices in the United States has fallen over the past several decades. For compensation measures where there was still evidence of pass-through, the variations had essentially no material effect on price inflation in the most recent period. Heise *et al* (2022) find that the pass-through from wages to prices in the goods-producing sector in the United States has fallen since the early 2000s, and that this fall has been an important source of low inflation. Similarly, Ascari and Fosso (2021) find that the pass-through from US wages to inflation has fallen. A further recent study, by Bobeica *et al* (2019), concluded that the pass-through of labour costs to inflation in the euro area has been higher in periods of high inflation than in periods of low inflation.

The recent looser connection between domestic wages and prices is broadly consistent with the earlier hypothesis put forward by Gordon (1988), that past wages help little in predicting inflation. Further, it aligns well with the interpretation of Forbes (2019), that globalisation has caused a flattening of the relation between domestic slack and CPI inflation.

The paper proceeds as follows. Section 2 presents the econometric methodology. Section 3 presents the estimates of the pass-through from domestic labour cost variation to inflation, and Section 4 analyses the key factors driving the reduction in pass-throughs. Robustness results are presented in Section 5, and Section 6 concludes.

## 2. Econometric Methodology

To analyse the pass-through from domestic labour markets to inflation, we examine data from 21 advanced economies.[3] This selection was based exclusively on data availability on core inflation, earnings and unit labour costs. Data sources are listed in the Appendix.

We use the mean group estimation method for dynamic heterogeneous panels of Pesaran and Smith (1995) in estimating the expression

$$\Delta p_{i,t} = (\rho + \mu_{1i}) \cdot \Delta p_{i,t-1} + (\lambda + \mu_{2i}) \cdot \Delta c_{i,t} + \alpha_i + \varepsilon_{i,t}, \quad (1)$$

where $\Delta p_{i,t}$ denotes inflation and $\Delta c_{i,t}$ unit labour cost growth in country *i* and year *t*. The country-specific components of the slope coefficients (i.e. $\mu_{1i}$ and $\mu_{2i}$) have zero means and constant covariances; $\alpha_i$ are country fixed effects, and $\varepsilon_{i,t}$ is the error term. Our key interest lies in the estimates of $\lambda$, which captures average effects, and in the country-specific values $\lambda_i = \lambda + \mu_{2i}$. These capture the extent of short-term pass-through from domestic labour cost growth to inflation. High $\lambda$s would indicate a tight connection between domestic labour markets and local prices. Additionally, we also examine the estimates of $\lambda/(1-\rho)$, which reflect the average pass-through of domestic labour costs to prices in the long run.

---

[3] They are Austria, Australia, Belgium, Canada, Czechia, Estonia, the euro area, Finland, France, Germany, Ireland, Italy, Japan, Luxemburg, the Netherlands, Portugal, South Korea, Sweden, Switzerland, the United Kingdom and the United States.

The rationale for the choice of Pesaran and Smith's heterogeneous slope model is that this model is flexible enough to allow labour market pass-throughs to CPI inflation to differ between countries. This is key because the weight of domestic factors in goods and services' prices is bound to differ across countries. Further, when the slope coefficients vary across groups, dynamic panel models estimated with fixed effects, instrumental variables or GMM can produce inconsistent estimates.[4]

## 3. A Weakening Link

The results of the estimated equation (1) for the effects of changes in domestic unit labour costs on headline CPI inflation are presented in Table 1. We present full sample estimates, as well as decade-specific estimates. The coefficient for changes in unit labour costs is significantly positive in the 1980s, at 0.276, but falls over time to an insignificant value of -0.002 in the 2010s. In other words, domestic labour costs have become much less important as a driver of headline CPI inflation across countries. The sharpest decline occurs around the turn of the millennium.

**CPI vs Unit Labour Costs**     Table 1

D.V.: $\Delta \ln$ CPI

|  | full sample | 1980s | 1990s | 2000s | 2010s |
|---|---|---|---|---|---|
| lagged $\Delta \ln$ CPI | 0.507*** | 0.297*** | 0.357*** | -0.169* | 0.385*** |
|  | 0.041 | 0.072 | 0.071 | 0.088 | 0.060 |
| $\Delta \ln$ ULC | 0.124*** | 0.276*** | 0.247*** | 0.064 | -0.002 |
|  | 0.033 | 0.075 | 0.052 | 0.061 | 0.063 |
| Constant | 0.007*** | 0.011*** | 0.007*** | 0.023*** | 0.008*** |
|  | 0.001 | 0.003 | 0.001 | 0.003 | 0.001 |
| LT effect |  |  |  |  |  |
| $\Delta \ln$ ULC | 0.252*** | 0.394*** | 0.385*** | 0.055 | -0.003 |
|  | 0.070 | 0.114 | 0.091 | 0.052 | 0.102 |
| observations | 735 | 108 | 186 | 210 | 210 |
| number of countries | 21 | 14 | 21 | 21 | 21 |
| RMSE (σ) | 0.0125 | 0.0113 | 0.0096 | 0.0105 | 0.0081 |
| χ2 | 167.87*** | 30.73*** | 48.21*** | 4.83* | 41.79*** |
| Wald test *p-value* | 0.000 | 0.000 | 0.000 | 0.089 | 0.000 |

Note: Pesaran and Smith (1995) mean group estimation based on yearly data between 1980 and 2020. Robust standard errors are shown below coefficients. \*\*\*/\*\*/\* denote statistical significance at 1/5/10% confidence level.

Next, we turn to core CPI inflation, which excludes the most volatile components such as oil and food items. The results of equation (1) for the effects of changes in unit labour costs on core CPI inflation are shown in Table 2. What is clear, is that also for core CPI inflation the coefficient on changes in unit labour costs is significantly positive in the 1980s, at 0.246, dropping sharply after the turn of the millennium to reach an insignificant value of 0.024 in the 2010s. The evolution of the pass-through coefficient to core inflation is plotted in Figure 1, with the respective confidence intervals.

---

[4] On this point, see Pesaran and Smith (1995) and Pesaran, Shin and Smith (1999).

**Core CPI vs Unit Labour Costs**          Table 2

D.V.: Δ ln Core CPI

|  | full sample | 1980s | 1990s | 2000s | 2010s |
|---|---|---|---|---|---|
| lagged Δ ln Core CPI | 0.588*** | 0.319*** | 0.411*** | 0.081 | 0.278*** |
|  | 0.033 | 0.082 | 0.075 | 0.069 | 0.075 |
| Δ ln ULC | 0.143*** | 0.246*** | 0.280*** | 0.036 | 0.024 |
|  | 0.022 | 0.075 | 0.051 | 0.035 | 0.032 |
| Constant | 0.004*** | 0.018*** | 0.005*** | 0.014*** | 0.008*** |
|  | 0.001 | 0.005 | 0.001 | 0.002 | 0.001 |
| LT effect |  |  |  |  |  |
| Δ ln ULC | 0.348*** | 0.361*** | 0.476*** | 0.039 | 0.034 |
|  | 0.060 | 0.169 | 0.105 | 0.038 | 0.045 |
| observations | 714 | 100 | 168 | 210 | 210 |
| number of countries | 21 | 13 | 18 | 21 | 21 |
| RMSE (σ) | 0.0107 | 0.0111 | 0.0094 | 0.0093 | 0.0047 |
| χ2 | 368.99*** | 25.77*** | 60.90*** | 2.41 | 14.33*** |
| Wald test *p-value* | 0.000 | 0.000 | 0.000 | 0.299 | 0.001 |

Note: Pesaran and Smith (1995) mean group estimation based on yearly data between 1980 and 2020. Robust standard errors are shown below coefficients. ***/**/* denote statistical significance at 1/5/10% confidence level.

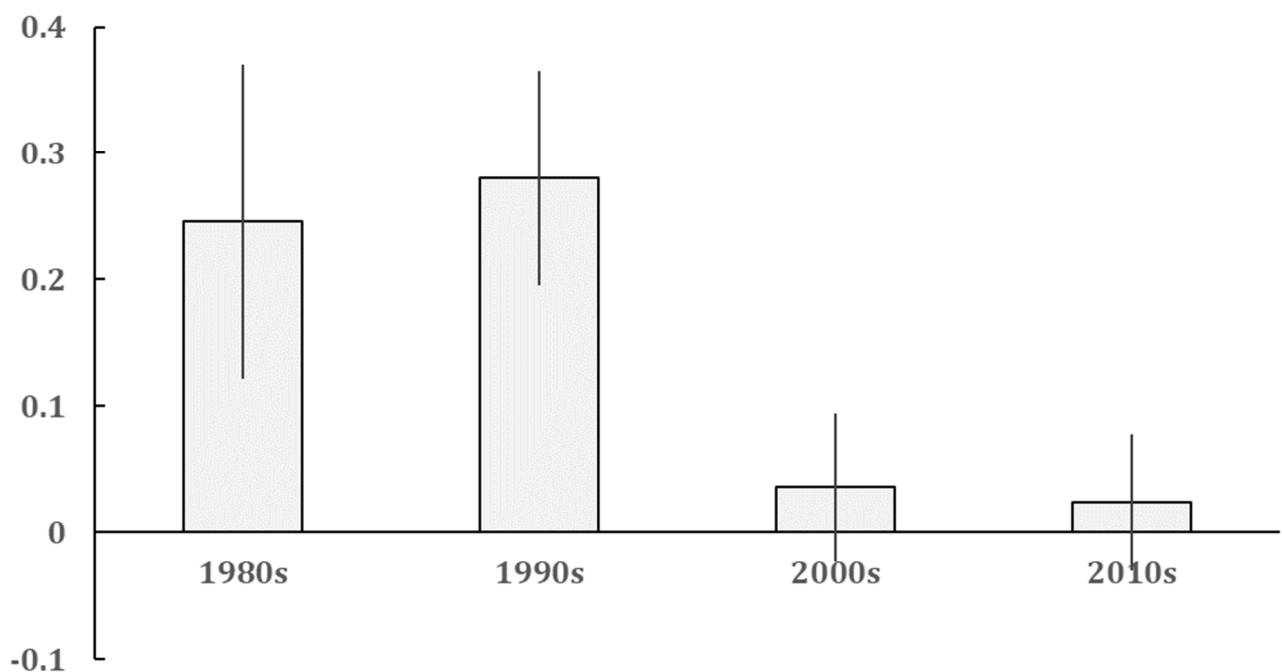

Fig. 1 – Labour market pass-through coefficients by decade (for core inflation, with 90% confidence intervals)

Importantly, the same result holds when the output gap is included in the regression as a control variable (see Table 3).[5] Appendix Table A1 shows that a similar conclusion is obtained if the unemployment gap is used as control instead. Further, we find that the result that the pass-through coefficient on changes in unit labour costs to core inflation falls sharply between the 1980s and the 2010s also goes through when replacing changes in unit labour costs by hourly earnings growth (see Table 4). What this indicates is that the relative decoupling is not driven by changing labour costs on the side of the employer.

---

[5] We start the regression of the model with the output gap in the 1990s rather than in the 1980s due to data availability issues.

**Models with output gap**      Table 3

D.V.: Δ ln Core CPI

|  | full sample | 1990s | 2000s | 2010s |
|---|---|---|---|---|
| lagged Δ ln Core CPI | 0.560*** | 0.442*** | 0.151 | 0.235*** |
|  | 0.039 | 0.090 | 0.097 | 0.087 |
| Δ ln ULC | 0.095*** | 0.200*** | -0.015 | 0.001 |
|  | 0.026 | 0.056 | 0.019 | 0.034 |
| output gap | 0.001*** | 0.002*** | 0.002 | 0.001*** |
|  | 0.000 | 0.001 | 0.000 | 0.000 |
| Constant | 0.007*** | 0.009*** | 0.010 | 0.011*** |
|  | 0.001 | 0.002 | 0.003 | 0.002 |
| LT effect |  |  |  |  |
| Δ ln ULC | 0.216*** | 0.357*** | -0.018 | 0.002 |
|  | 0.092 | 0.116 | 0.022 | 0.044 |
| observations | 655 | 154 | 209 | 210 |
| number of countries | 21 | 16 | 21 | 21 |
| RMSE (σ) | 0.0092 | 0.0092 | 0.0056 | 0.0043 |
| χ2 | 296.94*** | 46.49*** | 33.67*** | 18.07*** |
| Wald test *p-value* | 0.000 | 0.000 | 0.000 | 0.000 |

Note: Pesaran and Smith (1995) mean group estimation based on yearly data between 1980 and 2020. Robust standard errors are shown below coefficients. ***/**/* denote statistical significance at 1/5/10% confidence level.

**Core CPI vs Earnings**      Table 4

D.V.: Δ ln Core CPI

|  | full sample | 1980s | 1990s | 2000s | 2010s |
|---|---|---|---|---|---|
| lagged Δ ln Core CPI | 0.606*** | 0.398*** | 0.511*** | 0.114 | 0.309*** |
|  | 0.035 | 0.106 | 0.068 | 0.092 | 0.092 |
| Δ ln Earnings / hour | 0.136*** | 0.207** | 0.162*** | 0.030 | 0.060** |
|  | 0.026 | 0.081 | 0.037 | 0.041 | 0.029 |
| Constant | 0.003** | 0.013 | 0.004* | 0.010*** | 0.007*** |
|  | 0.001 | 0.009 | 0.002 | 0.004 | 0.002 |
| LT effect |  |  |  |  |  |
| Δ ln Earnings / hour | 0.347*** | 0.344** | 0.332*** | 0.034 | 0.087** |
|  | 0.074 | 0.147 | 0.088 | 0.046 | 0.044 |
| observations | 582 | 88 | 136 | 169 | 170 |
| number of countries | 17 | 11 | 14 | 17 | 17 |
| RMSE (σ) | 0.0109 | 0.0116 | 0.0104 | 0.0094 | 0.0046 |
| χ2 | 333.68*** | 20.68*** | 76.72*** | 2.07 | 15.62*** |
| Wald test *p-value* | 0.000 | 0.000 | 0.000 | 0.355 | 0.000 |

Note: Pesaran and Smith (1995) mean group estimation based on yearly data between 1980 and 2020. Robust standard errors are shown below coefficients. ***/**/* denote statistical significance at 1/5/10% confidence level.

## 4. Manufacturing EMEs Imports and ULC–CPI Pass-Throughs

The key question is what factors brought about this striking decoupling of domestic labour costs from CPI inflation in advanced economies?

One of the most salient developments during this time frame was the rapid integration of manufacturing EMEs, most notably from Asia as well as, Eastern Europe and Mexico, into the global economy and the rapid increase of their international trade flows that followed. Such integration exposed advanced economy producers to much more credible import competition from lower wage countries, with profound impacts on pricing.[6]

The manufacturing EME import penetration measure we use is defined as bilateral imports from the sum of the six countries used by Auer *et al* (2013) (i.e. China, India, Malaysia, the Philippines, Thailand and Mexico) divided by aggregate domestic demand in the advanced economy.[7],[8] The left side of Figure 2 shows the average evolution of this measure across advanced economies by decade. On the right side, we also show the same measure for the EM-10, which we define as the EM-6 plus Czechia, Hungary, Poland and Turkey. Even towards the end of the sample, all EM-10 countries had average wage levels that were below 40% of those in the United States.

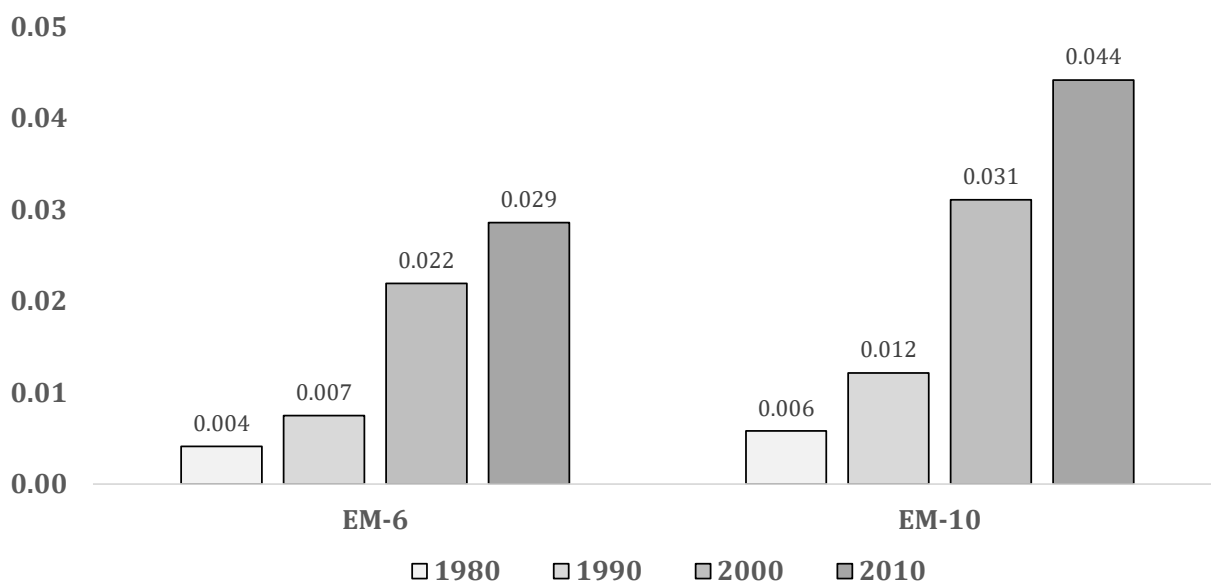

**Fig. 2 – Manufacturing EMEs' Import Penetration**
(median values for AEs)

Note: Imports from Manufacturing EMEs / Domestic Demand

Median manufacturing EME import penetration in advanced economies grew from just 0.4% in the 1980s to 2.9% in the 2010s for the EM-6, and from 0.6% to 4.4% if measured for the EM-10. The above increase in presence is likely to have impacted the pricing power of producers in the goods sector in advanced economies directly, through actual loss of market share, and indirectly, through increased market contestability.

To test whether the greater openness of advanced economies has impacted the pass-through from domestic labour costs to CPI inflation, we regress the estimated $\lambda_{i,t}$ for each advanced economy $i$ in our sample during decade $t$ against the respective KOF globalisation index (which is a broad measure of the degree of integration into the world economy), and against the EM-6 and EM-10 import penetration shares, according to

$$\lambda_{i,t} = \gamma \cdot globalisation_{i,t} + \alpha_i + \varepsilon_{i,t}, \qquad (2)$$

where $globalisation_{i,t}$ denotes the KOF globalisation index (or the EM-6 and EM-10 import penetration shares) in country $i$ and decade $t$. The results of this exercise can be seen in Table 5. In all six specifications, greater openness (corresponding to higher indices or shares) is associated with significantly lower estimated $\lambda_i$s. The *p-values* of the variables of interest are always below 0.05.[9]

---

[6] Auer *et al* (2013) for instance conclude that producer prices in Europe decreased by 3% for each percentage point of the European market that was captured by six developing manufacturing exporters.
[7] Bilateral trade flows are taken from the UN Comtrade database.
[8] That is, output + (imports – exports).

**The degree of pass-through vs LWC import penetration**  Table 5

| D.V.: Estimated pass-through coefficient (ULC → CPI) | | | | | | |
|---|---|---|---|---|---|---|
| | (I) | (II) | (III) | (IV) | (V) | (VI) |
| constant | -0.099 | -0.255 | -0.304* | -0.495** | -0.427** | -0.520** |
| | 0.073 | 0.156 | 0.162 | 0.229 | 0.166 | 0.234 |
| ln (globalisation index) | -1.108*** | -1.779** | | | | |
| | 0.362 | 0.670 | | | | |
| ln (EM-6 import penetration) | | | -0.108*** | -0.153** | | |
| | | | 0.041 | 0.054 | | |
| ln (EM-10 import penetration) | | | | | -0.150*** | -0.174** |
| | | | | | 0.046 | 0.060 |
| observations | 58 | 58 | 58 | 58 | 58 | 58 |
| country fixed effects | no | yes | no | yes | no | yes |
| R2 | 0.177 | | 0.123 | | 0.203 | |
| R2 within | | 0.246 | | 0.197 | | 0.237 |
| R2 between | | 0.176 | | 0.005 | | 0.124 |

Note: Robust standard errors are shown below coefficients. \*\*\*/\*\*/\* denote statistical significance at 1/5/10% confidence level.

## 5. Robustness of Results: Machine Learning and Interaction Terms

### 5.1 – Regression Decision Tree

Another potential variable that could affect the extent of pass-through from domestic labour costs to core CPI inflation is the level of inflation. Widespread adoption of inflation targeting regimes in advanced economies has led to lower inflation, lower inflation expectations and generally a much better anchoring of inflation around central banks' targets. As it turns out, the average level of inflation in AEs is highly correlated with the degree of globalisation and import penetration from manufacturing EMEs. This can be seen in Figure 3, for all three alternative indicators of openness that we use.

---

[9] All KOF globalisation indices and EME import penetration factors that were used are listed in Table A2. The analysis in this section does not include Czechia, Estonia, Luxemburg and South Korea.

# Figure 3 - Correlates of CPI Inflation

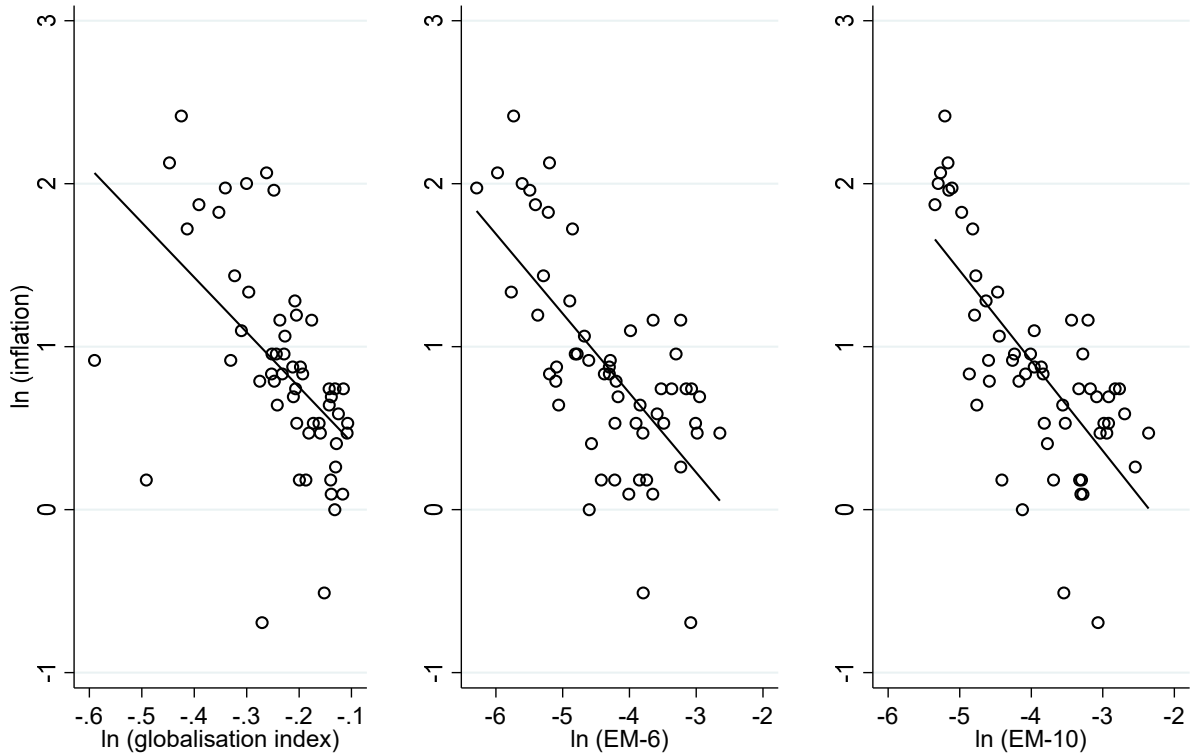

Note: Based on decade averages for advanced economies

Estimating which share of the long-term fall in estimated $\lambda$s is due to the inflation level per se, and which to increased competition from EMEs would in practice not be feasible by standard regression methods for the international panel in question. Regression estimates would not properly partition the contribution of individual factors if EME import penetration and inflation were included together as determinants of $\lambda$s. Since the number of observations for AEs is small (we have 58 estimates of $\lambda$s) one would not be able to get around the problem of multicollinearity, which is known to confound estimates of individual effects.[10] A much larger sample size would be required for multicollinearity not to be an issue.

Machine learning provides an alternative way for assessing the relative importance of EME import penetration and the inflation level for ULC–core CPI pass-throughs. More specifically we can use an entirely agnostic regression decision tree to fit our variable of interest (the estimated $\lambda$ for each country and decade). Essentially, the computer algorithm grows a decision tree mechanically using the reduction of mean square error as the splitting criterion at each node (see Breiman *et al* (1984)).[11] Once the decision tree has been grown, we can estimate the importance of each individual predictor. This step is based on the average difference between mean squared errors (MSEs) between the parent nodes and the total MSEs for the two splits.

---

[10] Indeed, due to the high correlation, EME import penetration might even be interpreted as an instrument for the inflation level.
[11] The regression tree method finds that the optimal first split of the data is based on whether EM-6 import penetration was above or below 0.00765 (for the case when EM-6 is used). It then continues with further splits of the data for the subsequent nodes.

Figure 4 shows predictor importance factors derived from individual regression decision trees. In one case, the algorithm to predict estimated $\lambda$s is fed with EM-6 import penetration and average CPI inflation over 10 years, and in the other with EM-10 import penetration and average inflation. In both cases, the metric indicates that EME import penetration is three or more times as important as the level of inflation in predicting the extent of pass-through of labour costs to CPI inflation.

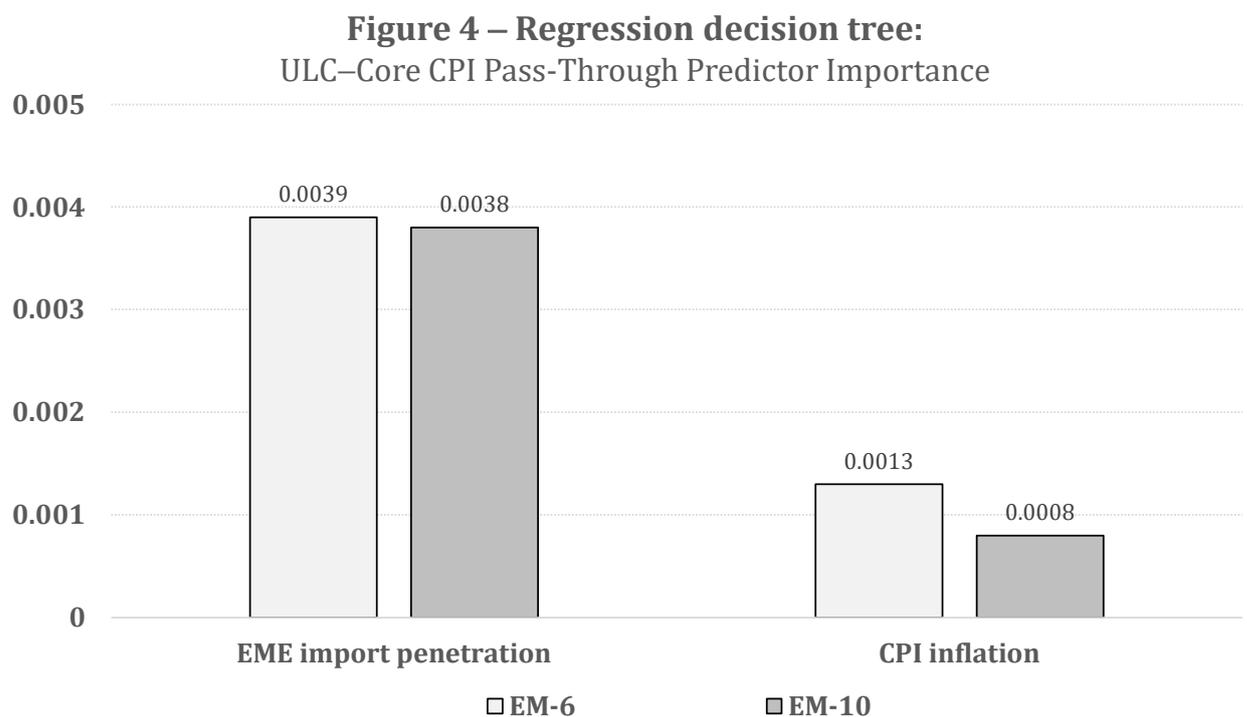

**Figure 4 – Regression decision tree:**
ULC–Core CPI Pass-Through Predictor Importance

We refine the above non-parametric analysis further, by examining how predicted domestic ULC–CPI pass-throughs vary with inflation and trade openness (here EM-10 import penetration). To produce robust results, we rely on an average prediction over 1,000 regression trees. Each tree is generated after randomized resampling, where 1/3 of the observations is discarded before a new regression tree is grown (i.e. we use "bagging"). The results which are shown in Figure 5 show that the pass-through rises non-linearly with the level of inflation. While it is generally negligible for inflation levels below 2% in more open economies, it jumps to about 20% for average inflation levels above 3%.[12] For more closed economies, the domestic ULC–CPI pass-through is typically very high, hovering around 40%.

---

[12] The average EM-10 import penetration index in the sample is 0.025. In the left panel we take 0.005 for a low trade openness economy and 0.045 for a high trade openness economy. The 3-D graph shows the variation for the full range.

As for the effects of trade openness itself, what the exercise reveals is that the pass-through drops precipitously after the initial opening (i.e. when EM-10 import penetration reaches the equivalent of close to 0.01 of domestic demand), irrespective of the average level of inflation. Beyond an import penetration of 0.02, further opening has little additional effect.[13] Note that with current EM-10 import penetration of respectively, 0.051 and 0.078, economies as the United States and Germany are well past this point of initial openness (see Table A2 in the Appendix). This suggests that market contestability in these is currently clearly high enough to keep the pass-through low.[14]

Domestic ULC – CPI pass-throughs: empirical sensitivity to determinants　　　　Figure 5

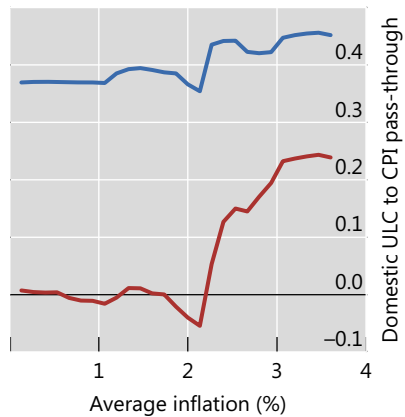
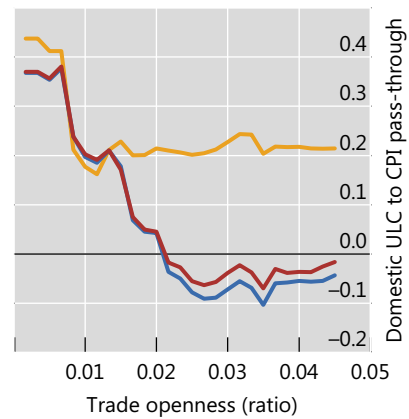
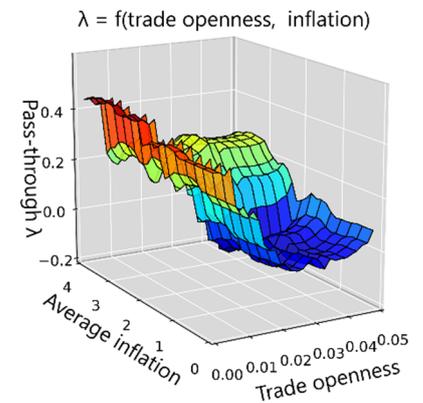

Estimates are based on a two-step exercise that relies on data from 21 advanced countries between 1980 and 2020. In the first step, the pass-through from domestic unit labour costs to CPI inflation for each country-decade are estimated based on the Pesaran-Smith mean group estimator. In the second step, an agnostic random forest model based on 1,000 regression trees is used to establish the link between these estimated pass-throughs with trade openness and average realized inflation (10-year average). Trade openness is defined as the ratio between imports from 10 major manufacturing based emerging market economies and the domestic component of aggregate demand in the receiving country.

## 5.2 – Interaction Terms

As a further robustness test, we also introduce an interaction term between unit labour cost variation and globalisation directly into the original panel regression,

$$\Delta p_{i,t} = (\rho + \mu_{1i}) \cdot \Delta p_{i,t-1} + (\lambda + \mu_{2i}) \cdot \Delta c_{i,t} + (\beta + \eta_i) \cdot globalisation_{i,t} +$$
$$+ (\gamma + \kappa_i) \cdot \Delta c_{i,t} \cdot globalisation_{i,t} + \alpha_i + \varepsilon_{i,t}, \qquad (3)$$

---

[13] The Code and data are available upon request.

where $globalisation_{i,t}$ denotes the KOF globalisation index in country $i$ and year $t$. This is an alternate approach to the second stage regressions presented above to determine whether globalisation affects the pass-through of unit labour costs to inflation. These estimation results are presented in Table 6 for both headline and core CPI inflation. The coefficient on the interaction term between unit labour cost growth and globalisation is significantly negative for both headline and core inflation, and of roughly similar magnitude. This implies that the pass-through of unit labour cost growth to both headline and core inflation is higher at lower levels of globalisation, which is consistent with the results presented above. We also find that the coefficient on unit labour costs remains positive and significant.

**Headline and core CPI vs Unit Labour Costs, with interaction term with globalisation**     Table 6

D.V.: $\Delta \ln$ CPI

|  | headline CPI inflation | core CPI inflation |
|---|---|---|
| lagged $\Delta \ln$ CPI | 0.327*** | 0.360*** |
|  | 0.043 | 0.049 |
| $\Delta \ln$ ULC | 1.996*** | 1.572*** |
|  | 0.461 | 0.447 |
| globalisation | -0.0003 | -0.001*** |
|  | 0.0002 | 0.0002 |
| globalisation* $\Delta \ln$ ULC | -0.023*** | -0.018*** |
|  | 0.006 | 0.005 |
| Constant | 0.031* | 0.056*** |
|  | 0.017 | 0.013 |
| observations | 651 | 651 |
| number of countries | 20 | 20 |
| RMSE (σ) | 0.0113 | 0.0098 |
| χ2 | 95.54 | 91.35 |
| Wald test *p-value* | 0.000 | 0.000 |

Note: Pesaran and Smith (1995) mean group estimation based on yearly data between 1980 and 2020. Robust standard errors are shown below coefficients. \*\*\*/\*\*/\* denote statistical significance at 1/5/10% confidence level.

Further, we also introduce an interaction term between unit labour cost growth and inflation lagged by two years, $\Delta p_{i,t-2}$, directly in the panel regression,

$$\Delta p_{i,t} = (\rho + \mu_{1i}) \cdot \Delta p_{i,t-1} + (\lambda + \mu_{2i}) \cdot \Delta c_{i,t} + (\tau + \vartheta_i)\Delta c_{i,t} \cdot \Delta p_{i,t-2} + \alpha_i + \varepsilon_{i,t}, \quad (4)$$

In the interaction term, inflation is lagged by two years in order to reduce endogeneity issues. This is an alternate approach to the second stage regressions presented above to determine whether the inflation environment affects the pass-through of unit labour costs to inflation.[15] The results of this exercise are presented in Table 7 for both headline and core CPI inflation. We find that the coefficient on the interaction term between unit labour cost growth and inflation lagged by two years is significantly positive for both headline and core inflation, and of similar magnitude. This suggests that the pass-through of unit labour cost growth to both headline and core inflation is larger in higher inflation environments.

---

[15] Our approach is similar to the one used in Jasova *et al* (2019) to determine whether the pass-through of exchange rate changes to inflation depends on the inflation environment.

**Headline and core CPI vs Unit Labour Costs, with interaction term with lagged inflation**      Table 7

D.V.: Δ ln CPI

|  | headline CPI inflation | core CPI inflation |
|---|---|---|
| lagged Δ ln CPI | 0.403*** | 0.476*** |
|  | 0.043 | 0.044 |
| Δ ln ULC | 0.057 | 0.042 |
|  | 0.038 | 0.031 |
| Δ ln CPI$_{t-2}$ * Δ ln ULC | 3.149** | 3.910*** |
|  | 1.500 | 1.368 |
| constant | 0.008*** | 0.006*** |
|  | 0.001 | 0.001 |
| observations | 702 | 698 |
| number of countries | 21 | 21 |
| RMSE (σ) | 0.0116 | 0.0100 |
| χ2 | 91.42 | 127.71 |
| Wald test *p-value* | 0.000 | 0.000 |

Note: Pesaran and Smith (1995) mean group estimation based on yearly data between 1980 and 2020. Robust standard errors are shown below coefficients. ***/**/* denote statistical significance at 1/5/10% confidence level.

Next, we introduce both the interaction terms, that is of unit labour cost growth with globalisation and with inflation lagged by two years, together in a single panel regression. This is to check which variable affects the pass-through of unit labour cost growth to inflation more,

$$\Delta p_{i,t} = (\rho + \mu_{1i}) \cdot \Delta p_{i,t-1} + (\lambda + \mu_{2i}) \cdot \Delta c_{i,t} + (\beta + \eta_i) \cdot globalisation_{i,t} +$$
$$+ (\gamma + \kappa_i) \cdot \Delta c_{i,t} \cdot globalisation_{i,t} + (\tau + \vartheta_i) \Delta c_{i,t} \cdot \Delta p_{i,t-2} + \alpha_i + \varepsilon_{i,t}, \quad (5)$$

The results are presented in Table 8. We find that the coefficient on the interaction term between unit labour cost growth and globalisation remains significantly positive for both headline and core inflation, and of similar magnitude as when the interaction term with globalisation is included on its own (see Table 6). By contrast, the coefficient on the interaction term between unit labour cost growth and inflation lagged by two years becomes statistically insignificant. This suggests that the pass-through of unit labour cost growth to both headline and core inflation is affected more by globalisation than by the inflation environment. This result is consistent with the result based on the regression decision trees analysis presented above.

**Headline and core CPI vs Unit Labour Costs, with interaction terms with globalisation and lagged inflation**  Table 8

D.V.: Δ ln CPI

|  | headline CPI inflation | core CPI inflation |
|---|---|---|
| lagged Δ ln CPI | 0.281*** | 0.315*** |
|  | 0.039 | 0.044 |
| Δ ln ULC | 2.379*** | 1.528*** |
|  | 0.5080 | 0.5720 |
| globalisation | -0.0003 | -0.0004 |
|  | 0.0002 | 0.0002 |
| globalisation* Δ ln ULC | -0.027*** | -0.017** |
|  | 0.0060 | 0.0070 |
| Δ ln CPI$_{t-2}$* Δ ln ULC | -1.485 | 0.298 |
|  | 1.180 | 1.705 |
| constant | 0.032* | 0.046*** |
|  | 0.017 | 0.017 |
| observations | 639 | 636 |
| number of countries | 20 | 20 |
| RMSE (σ) | 0.0107 | 0.0092 |
| χ2 | 95.62 | 69.88 |
| Wald test *p-value* | 0.000 | 0.000 |

Note: Pesaran and Smith (1995) mean group estimation based on yearly data between 1980 and 2020. Robust standard errors are shown below coefficients. \*\*\*/\*\*/\* denote statistical significance at 1/5/10% confidence level.

# 6. Concluding Remarks

This article examined the link between domestic labour markets and CPI inflation across advanced economies. It provides novel systematic cross-country evidence of the pass-through of domestic labour cost variations to CPI inflation, using dynamic panel estimation which allows for heterogeneity in coefficients across countries.

We find that the link between domestic unit labour cost growth and inflation has weakened considerably during recent decades. The short-run pass-through from domestic labour cost variation to core CPI inflation decreased from a significant value of 0.25 in the 1980s to just 0.02 in the 2010s. Similarly, the long-run pass-through fell from a significant value of 0.36 in the 1980s to 0.03 in the 2010s. Similar reductions are found for other measures of inflation. We show that the timing of the collapse in the pass-through coincides with a steep increase in import penetration from a group of 10 major manufacturing EMEs (low wage economies), which implies increased competition and market contestability in advanced economies.

Two alternative empirical tests reveal that globalisation has been the dominant factor explaining the decline in pass-through of domestic labour cost variations to inflation. While the lower level of inflation has also contributed to the decline, its contribution is overshadowed by that of greater economic openness. Overall, our results suggest that an excessive focus on domestic labour markets may have become less appropriate for gauging inflation pressures in a globalised economy.

# References

Ascari, G and L Fosso (2021): "On the international component of trend inflation and the flattening of the Phillips Curve", manuscript. University of Oxford.

Auer, R, K Degen and A Fischer (2013): "Low-wage import competition, inflationary pressure, and industry dynamics in Europe", *European Economic Review*, vol 59, 141–166.

Bobeica, E, Ciccarelli, M and I Vansteenkiste (2019): "The link between labor cost and price inflation in the euro area," Working Paper Series 2235, European Central Bank.

Breiman, L, J Friedman, R Olshen, and C Stone (1984). *Classification and Regression Trees*. Boca Raton, FL: CRC Press.

Dreher, A (2006): "Does globalization affect growth? Evidence from a new index of globalization", *Applied Economics*, 38(10), 1091–1110.

Forbes, K (2019): "Inflation dynamics: dead, dormant or determined abroad?" *Brookings Papers on Economic Activity*, Fall, pp. 257-338.

Gordon, R J (1988): "The role of wages in the inflation process", *American Economic Review*, pp. 276-283.

Gygli, S, Haelg, F, Potrafke, N and J-E Sturm (2019): "The KOF globalisation index – revisited", *Review of International Organizations*, 14(3), 543–574.

Jasova, M, Moessner, R and E Takats (2019): "Exchange rate pass-through: what has changed since the crisis?", *International Journal of Central Banking,* vol 15, no 3, 27–58.

Heise, S, Karahan, F and A Şahin (2022): "The missing inflation puzzle: The role of the wage-price pass-through", *Journal of Money, Credit and Banking*, vol 54, S1, 7-51.

Peneva, E and J Rudd (2017): "The passthrough of labor costs to price inflation", *Journal of Money, Credit and Banking,* 49(8), 1777–1802.

Pesaran, M H and R Smith (1995): "Estimating long-run relationships from dynamic heterogeneous panels", *Journal of Econometrics*, vol. 68(1), 79–113.

Pesaran, M H, Y Shin and R P Smith (1999): "Pooled mean group estimation of dynamic heterogeneous panels", *Journal of the American Statistical Association*, vol 94, 621–634.

# Appendix: Data Sources

Our panel consists of 21 advanced economies (listed in footnote 3). We use data on unit labour costs and compensation per hour from the OECD. Data on CPI headline inflation is taken from Datastream and the BIS. Data on CPI core inflation is from the OECD, national data and the BIS. Estimates of the output gap and the unemployment gap (defined as the unemployment rate minus the NAIRU) are taken from the OECD.

We also rely on data on import penetration from a group of 10 representative manufacturing based emerging market economies (EMEs) from UN Comtrade.[16] As a measure of globalisation we use the comprehensive KOF globalisation index, which quantifies the economic, social and political dimensions of globalisation (Dreher (2006); Gygli *et al* (2019)).

---

[16] They are China, Czechia, Hungary, India, Mexico, Malaysia, the Philippines, Poland, Thailand and Turkey.

**Models with unemployment gap**  Table A1

D.V.: Δ ln Core CPI

|  | full sample | 1990s | 2000s | 2010s |
|---|---|---|---|---|
| Δ ln Core CPI | 0.611*** | 0.463*** | 0.120 | 0.182*** |
|  | 0.035 | 0.084 | 0.082 | 0.066 |
| Δ ln ULC | 0.085*** | 0.184*** | -0.048** | 0.028 |
|  | 0.026 | 0.052 | 0.019 | 0.032 |
| unemployment gap | -0.003*** | -0.002* | -0.007*** | -0.002*** |
|  | 0.001 | 0.001 | 0.002 | 0.001 |
| Constant | 0.006*** | 0.009*** | 0.013*** | 0.010*** |
|  | 0.001 | 0.002 | 0.002 | 0.002 |
| LT effect |  |  |  |  |
| Δ ln ULC | 0.220*** | 0.344*** | -0.055* | 0.034 |
|  | 0.071 | 0.111 | 0.022 | 0.039 |
| observations | 665 | 164 | 209 | 210 |
| number of countries | 21 | 17 | 21 | 21 |
| RMSE (σ) | 0.0093 | 0.0090 | 0.0056 | 0.0044 |
| χ2 | 348.71*** | 45.95*** | 29.01*** | 15.89*** |
| Wald test *p-value* | 0.000 | 0.000 | 0.000 | 0.001 |

Note: Pesaran and Smith (1995) mean group estimation based on yearly data between 1980 and 2020. Robust standard errors are shown below coefficients. ***/**/* denote statistical significance at 1/5/10% confidence level.

**Table A2 – Globalisation Indices and EME Import Penetration**

| country | decade | KOF glob. index | EM-6 | EM-10 |
|---|---|---|---|---|
| AT | 1980 | 0.744 | 0.0031 | 0.0114 |
| AT | 1990 | 0.809 | 0.0062 | 0.0211 |
| AT | 2000 | 0.871 | 0.0154 | 0.0457 |
| AT | 2010 | 0.883 | 0.0277 | 0.0675 |
| AU | 1980 | 0.639 | 0.0055 | 0.0057 |
| AU | 1990 | 0.719 | 0.0137 | 0.0141 |
| AU | 2000 | 0.789 | 0.0395 | 0.0405 |
| AU | 2010 | 0.810 | 0.0524 | 0.0540 |
| BE | 2000 | 0.877 | 0.0429 | 0.0628 |
| BE | 2010 | 0.898 | 0.0306 | 0.0539 |
| CA | 1980 | 0.676 | 0.0045 | 0.0048 |
| CA | 1990 | 0.760 | 0.0150 | 0.0154 |
| CA | 2000 | 0.813 | 0.0345 | 0.0356 |
| CA | 2010 | 0.834 | 0.0505 | 0.0526 |
| CH | 1990 | 0.825 | 0.0055 | 0.0077 |
| CH | 2000 | 0.876 | 0.0100 | 0.0162 |
| CH | 2010 | 0.901 | 0.0268 | 0.0379 |
| DE | 1990 | 0.778 | 0.0084 | 0.0181 |
| DE | 2000 | 0.853 | 0.0224 | 0.0479 |
| DE | 2010 | 0.878 | 0.0395 | 0.0783 |
| FI | 1980 | 0.711 | 0.0019 | 0.0061 |
| FI | 1990 | 0.781 | 0.0061 | 0.0102 |
| FI | 2000 | 0.851 | 0.0202 | 0.0295 |
| FI | 2010 | 0.870 | 0.0237 | 0.0369 |
| FR | 1980 | 0.741 | 0.0037 | 0.0050 |
| FR | 1990 | 0.785 | 0.0064 | 0.0086 |
| FR | 2000 | 0.841 | 0.0147 | 0.0220 |
| FR | 2010 | 0.871 | 0.0260 | 0.0377 |
| GB | 1980 | 0.780 | 0.0041 | 0.0058 |
| GB | 1990 | 0.812 | 0.0075 | 0.0097 |
| GB | 2000 | 0.867 | 0.0215 | 0.0284 |
| GB | 2010 | 0.891 | 0.0294 | 0.0418 |
| IE | 1990 | 0.777 | 0.0135 | 0.0169 |
| IE | 2000 | 0.839 | 0.0261 | 0.0321 |
| IE | 2010 | 0.859 | 0.0225 | 0.0289 |
| IT | 1980 | 0.654 | 0.0032 | 0.0055 |
| IT | 1990 | 0.724 | 0.0051 | 0.0084 |
| IT | 2000 | 0.793 | 0.0126 | 0.0216 |
| IT | 2010 | 0.819 | 0.0213 | 0.0360 |
| JP | 1980 | 0.554 | 0.0100 | 0.0101 |
| JP | 1990 | 0.612 | 0.0120 | 0.0122 |
| JP | 2000 | 0.696 | 0.0297 | 0.0300 |
| JP | 2010 | 0.763 | 0.0459 | 0.0466 |
| NL | 1980 | 0.797 | 0.0093 | 0.0117 |
| NL | 1990 | 0.821 | 0.0136 | 0.0191 |
| NL | 2000 | 0.867 | 0.0465 | 0.0592 |
| NL | 2010 | 0.897 | 0.0710 | 0.0945 |
| PT | 1990 | 0.703 | 0.0054 | 0.0069 |
| PT | 2000 | 0.796 | 0.0081 | 0.0145 |
| PT | 2010 | 0.830 | 0.0147 | 0.0250 |
| SE | 1980 | 0.770 | 0.0025 | 0.0052 |
| SE | 1990 | 0.815 | 0.0046 | 0.0083 |
| SE | 2000 | 0.879 | 0.0104 | 0.0229 |
| SE | 2010 | 0.890 | 0.0182 | 0.0366 |
| US | 1980 | 0.661 | 0.0078 | 0.0081 |
| US | 1990 | 0.733 | 0.0186 | 0.0191 |
| US | 2000 | 0.784 | 0.0368 | 0.0377 |
| US | 2010 | 0.815 | 0.0493 | 0.0506 |
| country | decade | KOF glob. index | EM-6 | EM-10 |

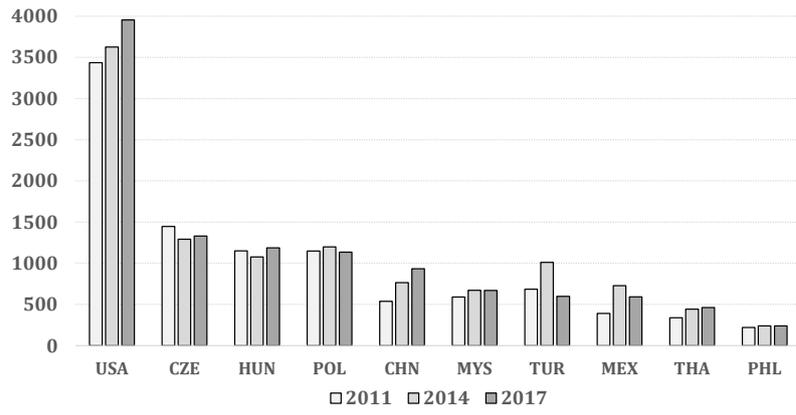

Figure A1 – Evolution of average monthly wage (USD)